\documentclass[aps,prl,twocolumn,showpacs,amsmath,amssymb,superscriptaddress]{revtex4-1}
\usepackage{graphicx}
\usepackage{dcolumn}
\usepackage{bm}
\usepackage{subfigure}
\newcommand{\qo}[1]{``#1''}                               
\newcommand{\ket}[1]{|#1\rangle}                      
\begin{document}
\title{Violation of Leggett-type inequalities in the spin-orbit degrees of freedom of a single photon}
\author{Filippo Cardano}
\affiliation{Dipartimento di Fisica, Universit\`{a} di Napoli Federico II, Complesso Universitario di Monte Sant'Angelo, Napoli, Italy}
\author{Ebrahim Karimi}
\altaffiliation[Current address: ]{Department of Physics, University of Ottawa, 150 Louis Pasteur, Ottawa, Ontario, K1N 6N5 Canada}
\email{\\ Corresponding author: ekarimi@uottawa.ca}
\affiliation{Dipartimento di Fisica, Universit\`{a} di Napoli Federico II, Complesso Universitario di Monte Sant'Angelo, Napoli, Italy}
\author{Lorenzo Marrucci}
\affiliation{Dipartimento di Fisica, Universit\`{a} di Napoli Federico II, Complesso Universitario di Monte Sant'Angelo, Napoli, Italy}
\affiliation{CNR-SPIN, Complesso Universitario di Monte Sant'Angelo, Napoli, Italy}
\author{Corrado de Lisio}
\affiliation{Dipartimento di Fisica, Universit\`{a} di Napoli Federico II, Complesso Universitario di Monte Sant'Angelo, Napoli, Italy}
\affiliation{CNR-SPIN, Complesso Universitario di Monte Sant'Angelo, Napoli, Italy}
\author{Enrico Santamato}
\affiliation{Dipartimento di Fisica, Universit\`{a} di Napoli Federico II, Complesso Universitario di Monte Sant'Angelo, Napoli, Italy}
\affiliation{Consorzio Nazionale Interuniversitario per le Scienze Fisiche della Materia, Napoli}
\begin{abstract}
We report the experimental violation of Leggett-type inequalities for a hybrid entangled state of spin and orbital angular momentum of a single photon. These inequalities give a physical criterion to verify the possible validity of a class of hidden-variable theories, originally named \qo{crypto non-local}, that are not excluded by the violation of Bell-type inequalities. In our case, the tested theories assume the existence of hidden variables associated with independent degrees of freedom of the same particle, while admitting the possibility of an influence between the two measurements, i.e. the so-called contextuality of observables. We observe a violation the Leggett inequalities for a range of experimental inputs, with a maximum violation of seven standard deviations, thus ruling out this class of hidden variable models with a high confidence.
\end{abstract}
\maketitle
\section*{Introduction}
Following Einstein, Podolsky and Rosen (EPR), a physical theory is considered \textit{complete} when there is a one-to-one correspondence between its elements and what are called the \textit{elements of reality} of the physical system that the theory aims to depict. Starting from the locality assumption, in the famous EPR paradox \cite{Eins35} it is argued that there exist physical systems, based on entanglement between different degrees of freedom, which are characterized  by elements of reality that do not have counterparts in the quantum theory. In this sense, therefore, quantum mechanics was said to be an incomplete theory. Following the seminal \textit{gedanken} experiment proposed by EPR, many scientists have tried to look for different theories, by introducing additional variables to the quantum ones, that recover the lack of knowledge (incompleteness) of the quantum theory; these additional variables are known as \textit{hidden variables}, because there is no known way to detect them experimentally~\cite{Bohm52v1,Bell64,Bell66}. The experimental violation of Bell inequalities has ruled out all possible forms of \textit{local hidden variable theories}, thus proving the impossibility of building up a theory that could recover both locality and so-called \qo{realism} at the same time~\cite{Aspe82,Clau69}. In 2003, A.~J.~Leggett formulated a new incompatibility theorem between quantum mechanics and a subclass of \textit{non-local} hidden variable theories, named \qo{crypto non-local}~\cite{Legg03}. In recent years, this (class of) non-local realism has been investigated in several experiments. The associated inequalities, named Leggett's inequalities, were also found to be experimentally violated, thus ruling out also this class of non-local hidden variables theories~\cite{Grob07,Bran07,Pate07,Bran08,Rome10}.

In all of the hitherto addressed experiments, the same degree of freedom of two entangled individual parties, for example the polarization of two entangled photons generated by spontaneous parametric down conversion, was used to test the Leggett inequalities. However, entanglement is not limited to a many particle system: separate degrees of freedom of a same particle can be used to generate a single particle entangled state. In such state, two (or more) observables of the particle may have a fully undetermined random value, but the overall quantum state of the particle is pure, thus implying the presence of strong correlations between the involved observables. Such correlations may then be tested against the predictions of hidden variable models. The Bell inequalities (or, alternatively, the tests of the Kochen-Specker kind), in this case, probe the so-called \qo{non-contextuality} of the observables, i.e. the assumption that the value of each observable as determined by the hidden values cannot be affected by the choice of measurement settings in place for other independent observables~\cite{Merm93}. On the other hand, the Leggett inequality goes beyond this, by testing the possible validity of a class of hidden-variable models, independently of the assumption for non-contextuality of the observables. Following Leggett, we might call this class of models \qo{crypto-contextual} models. A violation of Leggett's inequalities, therefore, negates the possible existence, before the act of measurement, of any elements of reality of the specific kind postulated in such crypto-contextual models.

In this article, we test a set of Leggett inequalities for single photons prepared in entangled states of the following two observables: spin angular momentum (SAM) and orbital angular momentum (OAM). SAM and OAM are respectively associated with the polarization and the wave front of an optical field and hence are fully independent observables in the paraxial limit~\cite{Allen92,Moli07,Marr11}. In quantum optics, SAM lives in a two-dimensional (2D) Hilbert space, while OAM corresponds to an unbounded, infinite-dimensional Hilbert space. However, we will use only a 2D subspace of OAM, thus mimicking a second spin state. In order to generate the single photon SAM-OAM entangled state, we exploited the spin-to-orbital coupling taking place in a suitably-patterned liquid crystal cell, named $q$-plate~\cite{Marr06,Marr11}. Similar applications of such device for quantum information experiments have been demonstrated in a number of recent works~\cite{Naga09prl,Naga09nphot,Kari10,Naga10,Damb12}. As we shall see, our experimental data for the SAM-OAM correlations were found to be in agreement with the quantum predictions and violating the Leggett's bound, with a maximum violation of seven standard deviations. This rules out with high confidence the possible validity of crypto-contextual hidden variable models for the single-photon SAM-OAM hybrid correlations.

\begin{figure}[t]
\centering
\includegraphics[width=9cm]{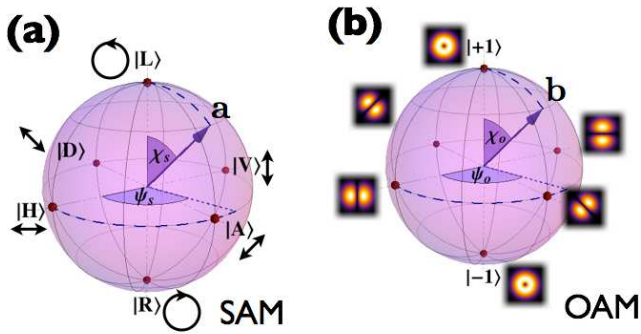}
\caption{(Color online) Geometric representation of the Hilbert spaces corresponding to the following single-photon observables: (a) SAM and (b) OAM, the latter restricted to a 2D spin-like subspace. North and south poles of the SAM Poincar\'e sphere correspond to left- and right-handed circularly polarized photons, respectively. Linearly polarized states are on the sphere equator. The OAM subspace is restricted to the eigenvalues $m=\pm1$. North and south poles of the corresponding sphere then represent $m=+1$ and $m=-1$, corresponding to the Laguerre-Gauss modes LG$_{0,1}$ and LG$_{0,-1}$, respectively (with the radial number $p=0$). The equator of the OAM Poincar\'e sphere corresponds to Hermite-Gauss modes, as given by a balanced superposition of the LG$_{0,1}$ and LG$_{0,-1}$ modes. The small insets in panel (b) show the intensity distribution patterns for some of these OAM modes.}
\label{Poincare}
\end{figure}
%
\section*{Leggett Inequalities}
In Leggett's model, we consider a physical system consisting of two 2D subsystems X and Y and two corresponding observables. In the original paper, the SAM (polarization) of two correlated photons was used, while in this article we take X to be the SAM and Y a 2D subspace of OAM of a single photon. For each of these observables, two observers Alice and Bob perform projective measurements specified by unit vectors ${\mathbf a}$ and ${\mathbf b}$ on X and Y, respectively. These vectors single out a state in the two-dimensional space and can be hence represented as unit vectors in the Poincar\'e sphere, as shown in Fig.\ \ref{Poincare}. The result $x$ ($y$) obtained by Alice (Bob) is a dichotomous variable, taking the possible values $+1$ and $-1$, where $+1$ is the outcome when the system is detected in state ${\mathbf a}$ ($\mathbf b$), and $-1$ when it is found to be orthogonal to state ${\mathbf a}$ ($\mathbf b$). In the crypto-contextual models, these possible outcomes are pre-determined (either deterministically or stochastically) by a set of hidden variables $\lambda$, defined in a domain $\Omega$. Due to our lack of knowledge about these hidden variables, all measured quantities result from an average in the hidden variable domain, weighted by the probability distribution $\rho(\lambda)$ that the system is in state $\lambda$.

In the original model proposed by Leggett \cite{Legg03}, the hidden variables $\lambda$ are defined by assigning a specific \qo{hidden} spin-like state $\mathbf{u}$ to X and $\mathbf{v}$ to Y, where these vectors live in the corresponding Poincar\'e spheres. This is done even if the overall quantum state is prepared in an entangled state for which the individual subsystems X and Y would have no definite quantum state. The expectation values for the observables X and Y with measurement settings $\mathbf{a}$ and $\mathbf{b}$, respectively, are then assumed to be given by the following expressions:
\begin{eqnarray}\label{avexy}
	\langle x \rangle_{\lambda} &=&\sum_x x\, P(x|\mathbf a,\mathbf b,\lambda) = \mathbf{u} \cdot \mathbf{a},\cr\cr
	\langle y \rangle_{\lambda} &=&\sum_y y\, P(y|\mathbf a,\mathbf b,\lambda) = \mathbf{v} \cdot \mathbf{b},
\end{eqnarray}
where $\lambda=(\mathbf{u},\mathbf{v})$. The correlation $C(\mathbf a,\mathbf b)$ between measurements performed by Alice and Bob are given by the expression
\begin{equation}\label{hvcorr}
	C(\mathbf a,\mathbf b)=\int_\Omega d\lambda\,\rho(\lambda)\left(\sum_{xy}xy\,P(x,y|\mathbf a,\mathbf b,\lambda)\right),
\end{equation}
where $P(x,y|{\mathbf a},{\mathbf b},\lambda)$ is a joint probability that the outcome of Alice and Bob projective measurement on ${\mathbf a}$ and ${\mathbf b}$ are $x$ and $y$, respectively. The model allows for contextuality of the two observables X and Y, because this joint probability is in general non-separable, i.e., $P(x,y|\mathbf a,\mathbf b,\lambda)\neq P(x|\mathbf a,\lambda)P(y|\mathbf b,\lambda)$. This in particular implies that each individual measurement outcome $x$ of observable $X$ may in general depend on the observable $Y$ settings ${\mathbf b}$, and possibly even on its simultaneous outcome $y$, and vice versa. However, it should be noted that Eqs.\ (\ref{avexy}) do imply a \qo{non--signalling} condition, so that each average value of a given observable is taken to be independent of the measurement settings of the other observable.

For this model, Branciard \textit{et al.} derived a simplified version of Leggett-type inequalities~\cite{Bran08}, which we will adopt here for our tests. This inequality involves three measurements on X along the vectors $\mathbf{a}_i$ and six on Y, along the vectors $\mathbf{b}_i$ and $\mathbf{b}'_i$, where $i=\{1,2,3\}$, with the following constraints: the three vector pairs $\mathbf{b}_i,\mathbf{b}'_i$ form a same angle $\phi$, their differences $\mathbf{b}_i-\mathbf{b}'_i$ must be three mutually orthogonal vectors, and their sums $\mathbf{b}_i+\mathbf{b}'_i$ must be respectively parallel to the $\mathbf{a}_i$. When these conditions are satisfied, Leggett's model gives rise to the following inequality~\cite{Bran08}:
\begin{eqnarray}\label{leggettinequality}
	E_3(\phi)&=&\frac{1}{3}\sum_{i=1}^3 \left |C(\mathbf {a_i},\mathbf {b_i})+C(\mathbf {a_i},\mathbf {b'_i})\right|\cr
	&\leq&2-\frac{2}{3}\left| \sin{\frac{\phi}{2}} \right|=L_3(\phi).
\end{eqnarray}
Quantum mechanics, on the other hand, predicts a violation of this inequality. Indeed, if the system is prepared in a maximally entangled state of the observables X and Y, such as for example $\ket{\Phi^+}=(\ket{+1}_X\ket{-1}_Y+\ket{-1}_X\ket{+1}_Y)/\sqrt{2}$, the correlation coefficients predicted by quantum mechanics are given by
\begin{equation}\label{QMcorr}
	C({\mathbf a},{\mathbf b})^{\tiny\hbox{QM}}=-{\mathbf a}\cdot{\mathbf b}=- \cos \left(\phi/2\right),
\end{equation}
from which we obtain
\begin{equation}
E_3^{\tiny\hbox{QM}}(\phi)=2|\cos{\phi/2}|.
\end{equation}
This function $E_3^{\tiny\hbox{QM}}(\phi)$ is above the Leggett bound of $\L_3(\phi)$ for a wide range of values of the angle $\phi$, as we will show further below.
\section*{Experiment}
\begin{figure}[h]
\centering
\includegraphics[width=9cm]{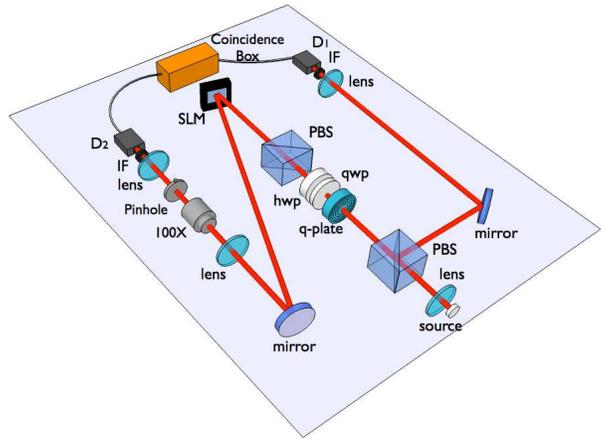}
\caption{(Color online) Experimental apparatus used for performing the Leggett test in the SAM-OAM Hilbert space of a single-photon. See text for a detailed explanation of the setup workings. Legend: PBS - polarizing beam splitter; hwp - half-wave plate; qwp - quarter-wave plate; SLM - spatial light modulator; 100X - microscope objective; IF - interference filters (10 nm bandwidth); D$_i$ - single-photon detectors.}
\label{setup}
\end{figure}

The layout of our experimental setup is shown in Fig.~\ref{setup}. The two-photon source (not shown in the figure) is based on a $\beta$-barium borate nonlinear crystal cut for degenerate collinear type-II phase matching, pumped by a linearly polarized beam at a wavelength of $397.5$ nm and at 100 mW of average power. The latter was obtained from the second harmonic of a Ti:Sapphire pulsed laser beam with a 100 fs pulse duration and 82 MHz repetition rate at the fundamental wavelength of 795 nm. For collinear type-II phase matching, the photon pair generated via spontaneous parametric down conversion is in the product state of $\ket{H}\ket{V}$, where $H$ and $V$ denote horizontal and vertical linear polarizations, respectively. The two photons are then separated by a polarizing beam splitter (PBS). The $V$-polarized photon was coupled directly to an avalanche single photon detector (D$_1$) by an appropriate set of lenses and mirrors and was used as trigger. The $H$-polarized photon, transmitted by the PBS, is used to perform the SAM-OAM measurements, by detecting the coincidences with the trigger photon, so as to operate in a heralded-single-photon quantum regime.

A tuned $q$-plate with topological charge $q=1/2$ was used to prepare the spin-orbit state of the photon \cite{Marr06,Naga09prl}. In particular, for a circularly polarized input photon, the $q$-plate flips the polarization helicity and simultaneously imparts $\pm2q\hbar$ of OAM to the photon, with the $\pm$ sign depending on the input polarization helicity. In other words, the $q$-plate gives rise to the following photon transformation laws: $\ket{L}_{\pi}\ket{0}_o\rightarrow\ket{R}_\pi\ket{2q}_o$ and $\ket{R}_{\pi}\ket{0}_o\rightarrow\ket{L}_\pi\ket{-2q}_o$, where $\pi$ and $o$ label the SAM and OAM degrees of freedom, respectively, $L$ and $R$ stand for the left and right circular polarizations, and $\ket{m}_o$ denotes an OAM eigenstate with eigenvalue $m\hbar$, such as for example a Laguerre-Gauss (LG) mode with azimuthal index $m$. The radial mode, as for example denoted by the LG index $p$, is irrelevant for our purposes, and it can be assumed hereafter to be fixed to $p=0$, as determined by the final detection mode. Now, a $H$ polarization state can be written as a superposition of left and right circular polarizations, i.e. $\ket{H}_{\pi}=(\ket{L}_{\pi}+\ket{R}_{\pi})/\sqrt{2}$. Hence, the $H$-polarized photon transmitted by the PBS is transformed by the $q$-plate into the following maximally-entangled spin-orbit state:
\begin{equation}\label{qplate:state}
	\ket{H}_{\pi}\ket{0}_{o}\stackrel{\small\hbox{$q$-plate}}{\longrightarrow}\ket{\Phi^+}=\frac{1}{\sqrt2}(\ket{L}_{\pi}\ket{-1}_{o}+\ket{R}_{\pi}\ket{+1}_{o}).
\end{equation}
It is worth noticing that the final state $\ket{\Phi^{+}}$ generated by the $q$-plate belongs to a four-dimensional Hilbert space defined as direct product of the SAM space and the OAM 2D subspace with $m=\pm1$. Such Hilbert space is for example spanned by the four product states $\ket{L}_\pi\ket{+1}_o$, $\ket{L}_\pi\ket{-1}_o$, $\ket{R}_\pi\ket{+1}_o$,  and $\ket{R}_\pi\ket{-1}_o$.

The heralded photon prepared in the single-photon SAM-OAM entangled state $\ket{\Phi^+}$ was then sent to the detection apparatus, where its SAM and OAM values, corresponding to observables X and Y, were both measured. The projective measurement on the SAM state of the photon was singled out by means of a properly-oriented sequence of a half-wave plate, a quarter-wave plate, and a polarizer. The orientations of the two wave plates define the selected projection state ${\mathbf a}$ of the measurement, in the SAM Poincar\'e sphere (Fig. \ref{Poincare}). Then, the OAM measurement was achieved by diffraction on a spatial light modulator (SLM) followed by a spatial-filter system composed of a lens, a 100X microscope objective, and a pinhole having a 1 mm radius. The OAM projection state corresponding to each vector $\mathbf{b}$ in the OAM Poincar\'e sphere of $|m|=1$ was thus determined by the hologram pattern visualized on the SLM, as was computed statically by a computer-generated-holography technique \cite{Kar09,Kari10}. The spatial filter was used to select only the TEM$_{00}$ Gaussian component of the diffracted beam in the far-field zone. The selected photon, after both projections, was finally coupled to another avalanche single-photon detector D$_{2}$. The signals from the two detectors D$_1$ and D$_2$ were read out by a coincidence box and a digital counter.
\begin{figure}[h]
\centering
\includegraphics[width=8.5cm]{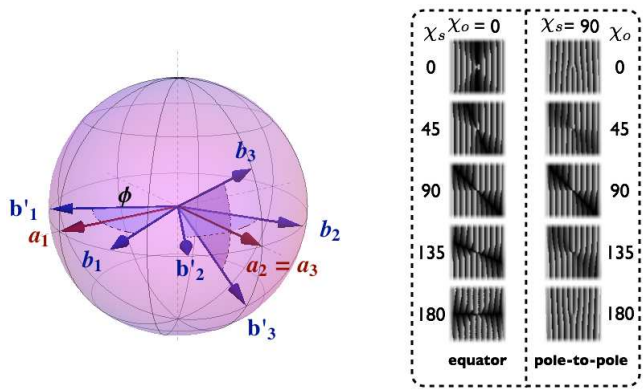}
\caption{(Color online) The set of state projections in the SAM and OAM states used to test the Leggett-type inequalities. The SAM (polarization) states to be measured were taken on the Poincar\'e sphere equator, i.e., $\mathbf{a}_1=\ket{H}$, $\mathbf{a}_2=\mathbf{a}_3=\ket{A}=(\ket{H}+\ket{V})/\sqrt{2}$. The OAM states $\mathbf{b}_1$ and $\mathbf{b'}_1$ were also taken along the equator line, at symmetrical azimuthal angles $\pm\phi/2$ relative to $\mathbf{a}_1$. Similarly, $\mathbf{b}_2$ and $\mathbf{b'}_2$ were taken along the equator line, at symmetrical angles $\pm\phi/2$ relative to $\mathbf{a}_2$. Finally, $\mathbf{b}_3$ and $\mathbf{b'}_3$ were taken along a meridian line, at symmetrical polar angles $\pm\phi/2$ relative to $\mathbf{a}_3=\mathbf{a}_2$. Examples of the computer-generated holograms needed to measure these OAM states are shown in the right inset, with $\chi_o$ and $\chi_s$ representing the polar and azimuthal angles (in degrees) on the OAM Poincar\'e sphere, respectively.}
\label{detectionsettings}
\end{figure}

The experimental correlation coefficients $C({\mathbf a},{\mathbf b})$ between the measurements of SAM and OAM are computed as
\begin{equation}\label{expcorr}
	C({\mathbf a},{\mathbf b})=\frac{N({\mathbf a},{\mathbf b})\!+\!N(-{\mathbf a},-{\mathbf b})\!-\!N({\mathbf a},-{\mathbf b})\!-\!N(-{\mathbf a},{\mathbf b})}{N({\mathbf a},{\mathbf b})\!+\!N(-{\mathbf a},-{\mathbf b})\!+\!N({\mathbf a},-{\mathbf b})\!+\!N(-{\mathbf a},{\mathbf b})},
\end{equation}
where $N({\mathbf a},{\mathbf b})$ are the experimental coincidence counts between the detectors D$_1$ and D$_2$ when the SAM and OAM projections are set to $\mathbf{a}$ and $\mathbf{b}$, respectively~\cite{Clau69}.

For the Leggett test, the adopted geometry of measurement settings $\mathbf{a}_i$ and $\mathbf{b}_i,\mathbf{b'}_i$ is shown in Fig.\ \ref{detectionsettings}, together with some representative holograms used to measure the OAM states.

\begin{figure}[t]
\centering
\includegraphics[width=7cm]{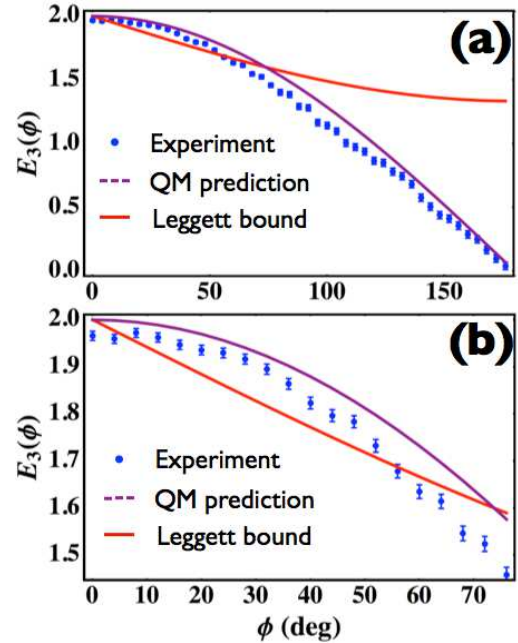}
\caption{(Color online) Experimental and theoretical values of Leggett's function $E_3(\phi)$. The blue points are the experimental data. Error bars are derived from Poissonian statistics on the coincidence counts and correspond to one standard deviation. The violet dashed line is the quantum mechanics prediction for the same function. The solid red line is the Leggett's bound $L_3(\phi)$. Panel (a) shows the entire range of measurement, $\phi\in[0,180^{\circ}]$. Panel (b) is a zoomed-in plot of the region in which a violation of the Leggett's bound is observed, i.e., $8^\circ \leq\phi\leq52^\circ$.}
\label{fig:e3data}
\end{figure}

Figure~\ref{fig:e3data} shows the experimental $E_3(\phi)$ data, as based on the measured correlation coefficients, for an angle $\phi$ varying within the range 0-180$^{\circ}$, in steps of $4^{\circ}$. The experimental data (blue points) are in good agreement with the predictions of quantum mechanics (violet dashed line), with only a small loss of visibility due to experimental imperfections. For a specific region, i.e., $8^\circ\leq\phi\leq52^\circ$, we obtained a violation of the Leggett bound. The maximum violation is found for $\phi=28^\circ$ and it is equal to $7.4\sigma$, where $\sigma$ is a standard deviation.

\section*{Conclusions}
Using Leggett-type inequalities, we have experimentally tested the possible validity of hidden-variable models for the measurement correlations between different degrees of freedom, namely spin and orbital angular momentum, in the case of a photon prepared in a single-particle entangled state of these two observables. The measured correlations agree with quantum predictions and hence violate the inequalities in a range of experimental parameters, thus showing with high confidence that for this physical system a wide class of deterministic models that preserve realism, even admitting a possible contextuality of the two observables, disagree with experimental results.

\section*{Acknowledgments}
We thank Miles J. Padgett and Sonja Franke-Arnold for valuable discussion. We acknowledge the financial support of the Future and Emerging Technologies (FET) programme within the Seventh Framework Programme of the European Commission, under FET-Open grant number 255914 - PHORBITECH. 


\begin{thebibliography}{99}

\bibitem{Eins35}
A.~Einstein, B.~Podolsky, and N.~Rosen, Phys.\ Rev.\  {\bf 47}, 777--780 (1935).

\bibitem{Bohm52v1}
D. Bohm, Phys. Rev. {\bf 85} 166--179 (1952).

\bibitem{Bell64}
J.~S.~Bell, Physics {\bf 1}, 195-- 200(1964).

\bibitem{Bell66}
J.~S.~Bell, Rev.\ Mod.\ Phys.\ {\bf 38}, 447--452 (1966).

\bibitem{Aspe82}
A.~Aspect, J.~Dalibard, and G.~Roger, Phys.\ Rev.\ Lett.\ {\bf 49}, 1804--1807 (1982).

\bibitem{Clau69}
J.~F.~Clauser, M.~A.~Horne, A.~Shimony, and R.~A.~Holt, Phys.\ Rev.\ Lett.\ {\bf 23}, 880--884 (1969).

\bibitem{Legg03}
A. J. Leggett, Foundations of Physics {\bf 33}, 1469 (2003).

\bibitem{Pate07}
T. Paterek, A. Fedrizzi, S. Gr{\"o}blacher, T. Jennewein, M {\.Z}ukowski, M Aspelmeyer, and A. Zeilinger, Phys. Rev. Lett. {\bf 99} 210406 (2007).

\bibitem{Grob07}
S. Gr{\"o}blacher, T. Paterek, R. Kaltenbaek, {\v C} Brukner, M. Zukowski, M. Aspelmeyer, and A. Zeilinger,  Nature {\bf 446}, 871--875 (2007).

\bibitem{Bran07}
C. Branciard, A. Ling, N. Gisin, C. Kurtsiefer, A. Lamas-Linares, and V. Scarani, Phys. Rev. Lett. {\bf 99}, 210407 (2007).

\bibitem{Bran08}
C. Branciard, N. Brunner, N. Gisin, C. Kurtsiefer, A. Lamas-Linares, A. Ling, and V. Scarani, Nat.\ Phys.\ {\bf 4}, 681 (2008).

\bibitem{Rome10}
J.~Romero, J.~Leach, B.~Jack, S.~M.~Barnett, M.~J.~Padgett, and S.~Franke-Arnold,  New Journal of Physics {\bf 12}, 123007 (2010).

\bibitem{Merm93}
N.~D.~Mermin, Rev.\ Mod.\ Phys.\ {\bf 65}, 803--815 (1993).

\bibitem{Allen92}
L.~Allen, M.~W.~Beijersbergen, R.~Spreeuw, and J.~P.~Woerdman, \pra {\bf 45} 8185 (1992).

\bibitem{Moli07}
G.~Molina-Terriza, J.~P.~Torres, L.~Torner, Nat.\ Phys.\ {\bf 3}, 305--310 (2007).

\bibitem{Marr11}
L. Marrucci, E. Karimi, S. Slussarenko, B. Piccirillo E. Santamato, E. Nagali, and F. Sciarrino, Journal of Optics {\bf 13} 064001 (2011).

\bibitem{Marr06}
L.~Marrucci, C.~Manzo, D.~Paparo, Phys.\ Rev.\ Lett.\ {\bf 96}, 163905 (2006).

\bibitem{Naga09prl}
E.~Nagali, F.~Sciarrino, F.~De Martini, L.~Marrucci, B.~Piccirillo, E.~Karimi, and E.~Santamato, Phys.\ Rev.\ Lett.\ {\bf 103}, 013601 (2009).

\bibitem{Naga09nphot}
E. Nagali, L. Sansoni, F. Sciarrino, F. De Martini, L. Marrucci, B. Piccirillo, E. Karimi, E. Santamato, Nature Phot.\ \textbf{3}, 720 (2009).

\bibitem{Kari10}
E. Karimi, J. Leach, S. Slussarenko, B. Piccirillo, L. Marrucci, L. Chen, W. She, S. Franke-Arnold, M.J. Padgett, and E. Santamato, Phys. Rev. A {\bf 82},  022115 (2010).

\bibitem{Naga10}
E. Nagali, D. Giovannini, L. Marrucci, S. Slussarenko, E. Santamato, F. Sciarrino, Phys.\ Rev.\ Lett.\ \textbf{105}, 073602 (2010).

\bibitem{Damb12}
V. D'Ambrosio, E. Nagali, S. P. Walborn, L. Aolita, S. Slussarenko, L. Marrucci, F. Sciarrino, Nature Comm.\ \textbf{3}, 961 (2012).

\bibitem{Kar09}
E.~Karimi, B.~Piccirillo, E.~Nagali, L.~Marrucci, and E.~Santamato, Appl.\ Phys.\ Lett.\, {\bf 94}, 231124 (2009). 

\end{thebibliography}
\end{document}